\documentclass[12pt]{JFM-FLM_Au}

\usepackage{color, soul}

\lefttitle{Ho Yin Ng and Emily S.C. Ching}
\righttitle{Journal of Fluid Mechanics}

\title{Eddy thermal diffusivity model and mean temperature profiles in turbulent vertical convection}

\author{Ho Yin Ng \and Emily S.C. Ching}

\affiliation{Department of Physics, The Chinese University of Hong Kong, Shatin, Hong Kong} 

\corresau{Emily S.C. Ching, \email{ching@phy.cuhk.edu.hk}}

\begin{document}
\maketitle

\begin{abstract}
In this paper, we propose a space-dependent eddy thermal diffusivity model for turbulent vertical natural convection in a fluid between two infinite vertical walls at different temperatures. Using this model, we derive analytical results for the mean temperature profile. Our results reveal that mean temperature profiles for different Rayleigh and Prandtl numbers are described by two universal scaling functions in the inner region next to the walls and the outer region near the centerline between the two walls  and the characteristic temperature scales in the inner and outer regions are expressed in terms of the two parameters of the model which determine the characteristic velocities for heat transfer in the two regions. We show that these results are in good agreement with direct numerical simulation data. \end{abstract}


\section{Introduction}
\label{intro}

Natural convective flows driven by temperature differences are ubiquitous in nature and engineering applications. To understand buoyancy-driven wall-bounded flows, it is common to study natural convection in a fluid confined between two vertical walls at different temperatures~\citep{Batchelor1954,ME1969,VN1999,BB2000,BHH2007,TSOPS2007,KH2012,NCO2013,Shishkina2016,HNVL2022} or
adjacent to a single heated vertical plate~\citep{Ostrach1953,Kuiken1968,Cheesewright1968,GC1979,RF1980,TN1988,KWAKN2021}. The state of these convective flows is determined by two control parameters, Rayleigh number ($Ra$) and Prandtl number ($\Pran$). Laminar vertical convection has been understood by analysis of steady-state boundary-layer equations~\citep{Ostrach1953,Kuiken1968,Shishkina2016} but a full understanding of turbulent vertical convection is still lacking. Knowledge of turbulent vertical convection is important for engineering applications such as ventilation in buildings and can shed light on ice-ocean interaction at near-vertical ice surfaces in a polar ocean~\citep{WW2008,HVL2023}. Physical quantities of interest include heat flux, wall shear stress, maximum mean vertical velocity and mean temperature and velocity profiles.  

A recent theoretical analysis by one of us~\citep{Ching2023} showed that for fluid confined between two infinite vertical walls at different temperatures, the Nusselt number ($Nu$) and shear Reynolds number ($Re_\tau$), which describe heat flux and wall shear stress, scale as $Ra^{1/3}$ in the high-$Ra$ limit. These theoretical results can well describe direct numerical simulation (DNS) data~\citep{HNVL2022}. The scaling $Nu \sim Ra^{1/3}$ is also consistent with experimental results for fluids with different $Pr$~\citep{Jakob1949,ME1969,TN1988} and the asymptotic law of heat transfer derived by~\citet{GC1979}  for turbulent natural convection next to a semi-infinite heated vertical plate. The analysis by~\cite{GC1979}  also yields the mean temperature and velocity profiles. By proposing scaling functions of temperature and velocity of certain characteristic scales of length, velocity and temperature in an inner layer next to the heated plate and a turbulent outer layer far away from the plate and matching them in an overlap layer, which is assumed to exist in the high-$Ra$ limit, they obtained an inverse cubic-root dependence on distance for the mean temperature and a cubic-root dependence for the mean velocity in the overlap layer. The result for the mean velocity deviates from both experimental and DNS data~\citep{VN1999,HH2005,SG2008}. Using a different temperature scale in the inner layer, a logarithmic mean temperature profile was obtained by \citet{HH2005}. The inverse cubic-root and the logarithmic mean temperature profiles have been shown to fit  experimental~\citep{Cheesewright1968,TN1988} and DNS data~\citep{VN1999,NCO2013} for air ($\Pran=0.709$) over different spatial regions but their validity for general values of $\Pran$ has not been tested. \citet{LJLJZ2023} studied the mean velocity and temperature profiles using models for turbulent heat flux and Reynolds stress but their models violate required boundary conditions at the vertical walls. 

In this paper, we propose a space-dependent eddy thermal diffusivity model for turbulent vertical convection in a fluid between two infinite vertical walls at different temperatures and use it to derive analytical results for the mean temperature profile. These analytical results reveal that mean temperature profiles for different Rayleigh and Prandtl numbers are described by two universal scaling functions in the inner region near the walls and the outer region near the centerline between the two walls. We validate our results using DNS data for $1 \le \Pran \le 100$ and $10^6 \le Ra \le 10^8$~\citep{HNVL2022} and show that the two scaling functions can better describe the data than results reported in previous studies.

\section{The problem}
\label{problem}

We consider a fluid confined between two infinite vertical walls separated by a distance $H$. The wall at the wall normal coordinate $x=0$ is kept at a temperature $T_h$ 
and the wall at $x=H$ at a lower temperature $T_c =T_h-\Delta T$.
With the Oberbeck-Boussinesq approximation, the equations governing the fluid motion are  $\nabla\cdot{\mathbf{u}}=0$ and
\begin{eqnarray}
    \frac{\partial \mathbf{u} }{\partial t}+{\mathbf{u}}\cdot\nabla {\mathbf{u}}&=&- \frac{1}{\rho} \mathbf{\nabla} p + \nu\nabla^2 {\mathbf{u}}+\alpha g(T-T_m)\hat{z}
    \label{Meqn}\\
 \frac{\partial T}{\partial t}+{\mathbf{u}}\cdot\nabla T&=&\kappa\nabla^2T
    \label{Teqn}
\end{eqnarray}
where $\mathbf{u}(x,y,z,t)=(u,v,w)$ is the velocity, $p(x,y,z,t)$ is the pressure,
$T(x,y,z,t)$ is the temperature, $\rho$ is the density of the fluid at the temperature at the centerline between the walls, $T_m=(T_h+T_c)/2$, $\alpha$, $\nu$ and $\kappa$ are the volume expansion coefficient, kinematic viscosity and thermal diffusivity of the fluid, respectively, $g$ is the acceleration due to gravity and
 $\hat{z}$ is a unit vector along
the vertical direction. The velocity field satisfies the no-slip boundary condition
at the two walls. The Rayleigh number and the Prandtl number are defined by
\begin{eqnarray}
Ra \equiv \frac{\alpha g  \Delta T H^3}{\nu \kappa}, \qquad \Pran \equiv \frac{\nu}{\kappa}
\label{RaPr}
\end{eqnarray}

The flow quantities are Reynolds decomposed into sums of time
averages and fluctuations such as $T(x,y,z,t)= \overline{T}(x,y,z)+T'(x,y,z,t)$, where an overbar denotes an average over time and primed symbols denote fluctuating quantities. 
As the vertical walls are infinite, all the mean flow quantities depend on $x$ only. This is also valid when periodic boundary conditions are imposed on the velocity and temperature in the spanwise ($y$) and streamwise ($z$) directions  as in DNS~\citep{VN1999,NCO2013,HNVL2022}. 
Taking the time
average of (\ref{Meqn}) and (\ref{Teqn}) leads to the mean
flow equations~\citep{VN1999}
\begin{eqnarray}
    \frac{d}{dx} \overline{u' w'} &=&\nu \frac{d^2}{dx^2} \overline{w} +\alpha g (\overline{T} - T_m)
   \label{MBE}\\
    \frac{d}{dx} \overline{ u'T'}&=&\kappa \frac{d^2}{dx^2} \overline{T}
    \label{MTE}
\end{eqnarray}
The  boundary conditions are
\begin{eqnarray}
  && \overline{w}(0)=0, \qquad \overline{w}(H/2)=0, \\ \label{BCw}
  && \overline{T}(0)= T_h,  \qquad  \overline{T}(H/2)=T_m 
\label{BCT}
\end{eqnarray}
and by symmetry, the mean profiles $\overline{w}(x)$ and $\overline{T}(x)$ are antisymmetric about $x=H/2$. 
A well-known challenge for solving $\overline{w}(x)$ and $\overline{T}(x)$ is that (\ref{MBE}) and (\ref{MTE}) are not closed. 
In this paper, we focus on finding $\overline{T}$ by solving (\ref{MTE}) with (\ref{BCT}) using a closure model for the eddy thermal diffusivity.

\section{Eddy thermal diffusivity and mean temperature profiles}
\subsection{Eddy thermal diffusivity model}
\label{Emodel}

Integrating~(\ref{MTE}) with respect to $x$, one obtains
\begin{equation}
 \overline{u'T'} - \kappa \frac{d\overline{T}}{dx} = - \kappa \frac{d \overline{T}}{dx} \bigg|_{x=0} =  \kappa Nu \frac{\Delta T}{H}
  \label{MTEI}
\end{equation} where $Nu$, being the ratio of the actual heat flux normal to the walls to the heat flux when there is only thermal conduction, is defined by 
\begin{equation}
Nu \equiv \frac{\overline{ u' T'} - \kappa d\overline{T} /dx}{\kappa \Delta T/H}  \equiv \frac{q}{\kappa \Delta T/H}
\label{Nuq}
\end{equation}
The heat flux is given by the product of $q$ and $\rho c$, where $c$ is the specific heat capacity of the fluid. 
We introduce a space-dependent function $K(x)$ for the eddy thermal diffusivity:
\begin{equation}
\overline{u'T'}  \equiv - K(x)  \frac{d\overline{T}}{dx}
\label{eddy}
\end{equation}
Then (\ref{MTEI}) with (\ref{eddy}) can be integrated to give~\citep{RF1980}
\begin{equation}
\frac{1}{\Delta T} [\overline{T}(x)-T_m]=  \frac{1}{2} - Nu  \int_0^{x/H} \frac{dy}{1+K(yH)/\kappa} ,
\label{meanT}
\end{equation}
and $Nu$ is obtained by using the boundary condition at $x=H/2$ in~(\ref{BCT}):
\begin{equation} 
Nu = \frac{1}{2} \left[ \int_0^{1/2}  \frac{dy}{1+K(yH)/\kappa} \right]^{-1}
\label{Nueq0}
\end{equation}
Thus, both $\overline{T}$ and $Nu$ can be determined if $K(x)$ is known.

The characteristic temperature scales in the inner region near the wall and the outer region near the centerline between the two walls should have different dependencies on $\nu$ and $\kappa$ as molecular diffusivities are expected to play a significant role in the inner region but not the outer region. As seen from (\ref{meanT}), this suggests that the eddy diffusivity $K(x)$ is governed by two independent parameters in the inner and outer regions. Thus, we model $K(x)$ separately in the inner and outer regions and in the intermediate region between them. Due to the no-slip and isothermal boundary conditions at the two vertical sidewalls, 
 $u'$, $T'$ and $\partial u'/\partial x=-(\partial v'/\partial y+ \partial w'/\partial z)$ vanish at $x=0$, and this leads to the vanishing of
 $\overline{u'T'}$ and its first- and second-order derivatives with respect to $x$ at $x=0$~\citep{RF1980,Ching2023}.  Since the temperature gradient at $x=0$, being proportional to $Nu$, is non-zero, 
 $K(x)$ and its first- and second-order derivatives should also vanish at $x=0$. Thus, we  model $K(x)/\nu$ by a cubic function in the inner region. In the outer region, we make use of the symmetry of the problem. Since $\overline{u'T'}$ is symmetric and  $\overline{T}$ is antisymmetric about the centerline $x=H/2$, $K(x)$ is symmetric about $x=H/2$ and attains a maximum value at $x=H/2$. This motivates us to model $K(x)/\nu$ by the simplest symmetric quadratic function with a peak at $x=H/2$ in the outer region. For the intermediate region, we choose a linear function that connects continuously to the other two regions as it is simple and the resulting $K(x)$ gives rise to a $Nu$ that scales as $Ra^{1/3}$ in the high-Ra limit in accord with the theoretical result~\citep{Ching2023}~(c.f. Sec.~\ref{valid}). Putting all these results together, we propose a three-layer model for $K(x)/\nu$:  
\begin{eqnarray}
\frac{K(x)}{\nu} = \begin{cases} A y^3, & 0 \le y \le y_1 \\
c_1 + c_2 y, &  y_1 < y < y_2  \\
C_m[1- b(1/2-y)^2], & y_2 \le y \le 1/2 
\end{cases}
\label{model} ,
\end{eqnarray}
where $y=x/H$. Near the walls, the turbulent heat flux is expected to be small compared to the conductive heat flux, or equivalently,  $K(x)$ is expected to be small compared to $\kappa$. Thus, a natural characteristic length scale for the inner region is given by
\begin{equation} 
 l_i = \frac{H}{(\Pran A)^{1/3}} ,
 \label{linner}
 \end{equation}
which is the position at which our proposed model of eddy thermal diffusivity is equal to $\kappa$, and we set the width of the inner region to be $2 l_i$, i.e., $y_1 H= 2 l_i$.
 We note that the equality of the eddy and molecular thermal diffusivities was used to define the thickness of the thermal boundary layer in vertical convection by~\citet{HNVL2022}.
We fix the width of the outer region to be $0.2 H$ with $y_2=0.3$ and take $b=4$. These values are guided by the DNS data~\citep{HNVL2022}. The constants $c_1$ and $c_2$ are fixed by the continuity of $K(x)/\nu$ at $x=y_1H$ and $x=y_2H$:
\begin{eqnarray}
c_1 = \frac{Ay_1^3 y_2-C_m\left[1-b(1/2-y_2)^2\right]y_1}{y_2-y_1}, \qquad c_2 = \frac{C_m\left[1-b(1/2-y_2)^2\right]- Ay_1^3}{y_2-y_1} \ \ 
	\end{eqnarray}
Hence, the model has two independent parameters: $A$ and $C_m$. We will show below that they determine the characteristic velocities for heat transfer in the inner and outer regions, respectively.

\subsection{Mean temperature profiles}

Substitute (\ref{model}) into (\ref{meanT}) and evaluate the integral, we obtain $\overline{T}(x)$:
 \begin{eqnarray}
 \frac{1}{\Delta T} [T_h - \overline{T}(yH)] &=& Nu \ I_1(y),  \qquad 0 \le y \le y_1 \label{analytic1}\\
\frac{1}{\Delta T} [\overline{T}(yH)-T_{m}] &=& \begin{cases} Nu \  [I_1(y_1)+I_2(y)], & y_1 < y < y_2 \\
 Nu \ I_3(y), \qquad y_2 \le y \le 1/2 \label{analytic2}
 \end{cases}
\end{eqnarray}
where 
\begin{equation}
Nu = \frac{1}{2[I_1(y_1)+ I_2(y_2) + I_3(y_2)]} ,
\label{Nueqn}
\end{equation}
and 
\begin{eqnarray}
\nonumber I_1(y) &\equiv&\frac{1}{\Pran^{\frac{1}{3}} A^{\frac{1}{3}}} \int_0^{Pr^{1/3} A^{1/3}y} \frac{dy' }{1+  y'^3} 
=   \frac{1}{3\Pran^{\frac{1}{3}} A^{\frac{1}{3}}} \left\{ \frac{1}{2} \log\left[\frac{(1+\Pran^\frac{1}{3} A^\frac{1}{3} y)^3}{1+\Pran A y^3}\right] \right. \nonumber  \\ 
&& \mbox{} \hspace{4cm} +\left. \sqrt{3}\arctan \left(\frac{2\Pran^\frac{1}{3} A^\frac{1}{3} y -1}{\sqrt{3}} \right) + \frac{\sqrt{3} \pi}{6} \right\} ,  \ \label{I1}\\
 I_2(y) &\equiv& \int_{y_1}^{y} \frac{dy' }{1+  Pr(c_1 + c_2 y')} =
\frac{1}{\Pran c_2} \log \left[ \frac{1+\Pran(c_1+ c_2y)}{1+Pr(c_1 + c_2 y_1)} \right] \label{I2}, \\
\nonumber I_3 (y) &\equiv&  \int_{y}^{1/2} \frac{dy' }{1+  \Pran C_m[1- b(1/2-y')^2]}  \\
 &=& \frac{1}{2B(1+C_m \Pran)} { \log \left| \frac{1 + B (1/2-y)}{1 - B (1/2-y)} \right|} , \ \  B= \sqrt{b C_m Pr/(1+C_m \Pran)} \qquad
\label{I3}
\end{eqnarray} 

We focus on the mean temperature profile in the inner and outer regions. The analytical results (\ref{analytic1}) and (\ref{analytic2}) reveal that $\overline{T}(x)$ is given  by two universal scaling functions of their respective characteristic temperature and length scales in these two regions, namely
\begin{eqnarray}
T_h - \overline{T}(x) &=& T_iF_i (x/l_i) \qquad \mbox{inner wall region} \label{Tinner}\\
\overline{T}(x) - T_m &=& T_o F_o(x/H) \qquad \mbox{outer centerline region} \label{Touter}
\end{eqnarray}
where the inner and outer scaling functions are
 \begin{eqnarray}
F_i(z) &=& \frac{1}{6}\log\left[\frac{(1+z)^3}{1+z^3}\right] + \frac{\sqrt{3}}{3} \arctan \left( \frac{2z -1}{\sqrt{3}}\right) + \frac{\sqrt{3}\pi}{18} \label{scalinginner} \\
 F_o(z) &=& \frac{1}{2\sqrt{b}} \log \left| \frac{1+\sqrt{b}(1/2-z)}{1-\sqrt{b}(1/2-z)} \right|  = \frac{1}{4} \log \left(\frac{1 -z}{z}\right), \label{scalingouter}
\end{eqnarray}
the inner and outer temperature scales are
\begin{eqnarray}
T_i &=& \frac{\Delta T Nu}{(\Pran A)^{1/3}}  \label{Tiscale} \\
T_o &= &\frac{\Delta T Nu}{\Pran C_m}  , 
\label{To}
\end{eqnarray}
and the characteristic length scales in the inner and outer regions are $l_i$ and $H$, respectively.
To obtain (\ref{Touter}) with (\ref{scalingouter}), we have used the approximation $\Pran C_m \gg 1$, which follows from the dominance of the turbulent heat flux over the conductive heat flux, $|\overline{u'T'}|  \gg \kappa |d \overline{T}/dx|$, at $x=H/2$ in the outer region for turbulent convective flows. Using (\ref{Nuq}) and (\ref{linner}), we can rewrite (\ref{Tiscale}) and (\ref{To}) as
\begin{equation}
V_i T_i = V_o T_o = q 
 \label{qqrelation}
\end{equation}
where 
\begin{eqnarray}
V_i &\equiv& \frac{\kappa}{l_i}  = \frac{\kappa}{H} (\Pran A)^{1/3} \label{Vi} \\
 V_o  &\equiv& \frac{K(H/2)}{H}  = \frac{\nu C_m}{H}
\label{Vo}
\end{eqnarray}
Thus, the two parameters $A$ and $C_m$ determine $V_i$ and $V_o$, the characteristic velocities for heat transfer in the inner and outer regions, respectively.

\section{Validation and discussions}
\label{valid}

\subsection{Validation}

\begin{figure}[h]
  \centerline{\includegraphics[scale=0.35]{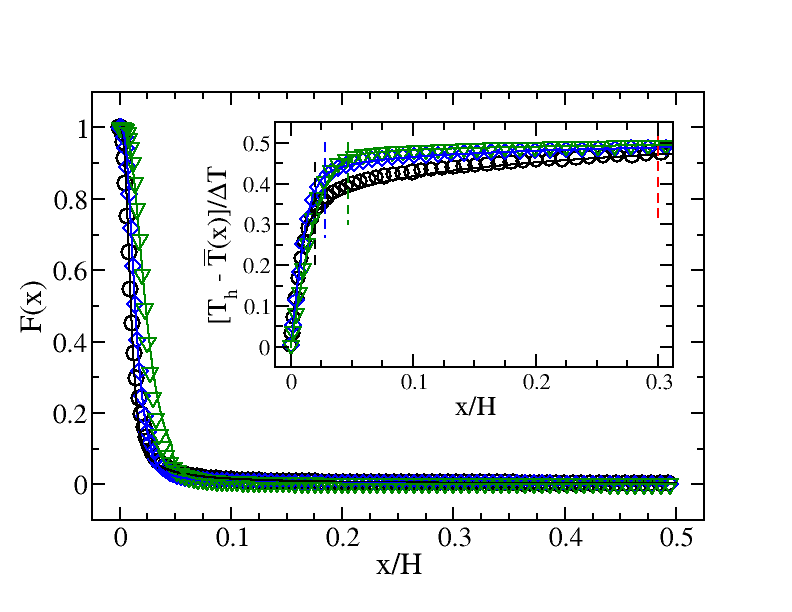}}
  \caption{ $F(x) = 1/[1+\Pran K(x)/\nu]$ vs $x$ at $Ra=10^8$ for $\Pran=1$~(black),
   $\Pran=10$~(blue) and $\Pran=100$~(green). The symbols are DNS data~\citep{HNVL2022} and the solid lines are evaluated using the eddy diffusivity model~(\ref{model}) with the values of $A$ and $C_m$ from table~\ref{tab:tab1}. In the inset, we compare the analytical result of the mean temperature profiles $\overline{T}(x)$ from the model (solid lines) with the DNS data.The dashed lines denote the positions of $y_1$ and $y_2=0.3$.} 
\label{fig:Nfig1}
\end{figure}

Using DNS data for $\Pran=1, 2, 5, 10$ and $100$ and $Ra$ ranging from $10^{6}$ to $10^{9}$~\citep{HNVL2022}, we evaluate $K(x)/\nu$, fit
it by $A (x/H)^3 $ in the region $0 \le x/H \le  2 l_i/H=2/(Pr A)^{1/3}$ using least-square method to obtain $A$ and extract $C_m$ directly as $C_m=K(H/2)/\nu$. 
The values of $A$ and $C_m$ are shown in table~\ref{tab:tab1}.
In determining $\overline{T}(x)$ and $Nu$, the relevant quantity is $F(x) = 1/[1+ \Pran K(x)/\nu]$ [see (\ref{meanT}) and (\ref{Nueq0})]. 
We compare $F(x)$ evaluated using the eddy diffusivity model (\ref{model}) and the DNS data in figure~\ref{fig:Nfig1}. Good agreement can be seen, which is further supported by the relatively small errors of the values of $Nu$ estimated using the model with respect to the DNS data~(see table~\ref{tab:tab1}).  Good agreement between the analytical result for $\overline{T}$ and the DNS data is also shown in figure~\ref{fig:Nfig1}. If $K(x)/\nu$ is modeled by $A(x/H)^3$ throughout the whole region~\citep{RF1980}, the estimated $Nu$ would be given by $1/[2 I_1(1/2)]$. As shown in table~\ref{tab:tab1}, the relative errors of these estimates of $Nu$ are substantially larger for $\Pran \le 5$.

\begin{table}
 \begin{center}
\def~{\hphantom{0}}
\begin{tabular}{ccrrrrr}
\hline
\Pran & $Ra$ & $A$ & \hspace{0.6cm} $C_m$  &  \hspace{0.1cm} $Nu$ (DNS)  &   \hspace{0.4cm} $Nu$ (model) & \hspace{0.4cm}  $1/[2I_1(1/2)]$ \\\hline
1 & $1 \times 10^6$ & 10478.14 & 30.77 & 6.59 & 7.22 (9.51\%) & 9.08 (37.70\%) \\
1 & $2 \times 10^6$ & 22731.58 & 41.78 & 8.30 & 9.06 (9.12\%) & 11.74 (41.36\%) \\
1 & $5 \times 10^6$ & 56711.28 & 60.42 & 11.31 & 11.92 (5.38\%) & 15.90 (40.66\%) \\
1 & $1 \times 10^7$ & 109126.23 & 80.68 & 13.81 & 14.63 (5.91\%) & 19.77 (43.18\%) \\
1 & $2 \times 10^7$ & 225663.87 & 109.19 & 17.10 & 18.34 (7.23\%) & 25.19 (47.28\%) \\
1 & $5 \times 10^7$ & 521909.56 & 168.07 & 22.93 & 24.40 (6.39\%) & 33.30 (45.21\%) \\
1 & $1 \times 10^8$ & 972659.79 & 221.83 & 28.19 & 29.91 (6.10\%) & 40.98 (45.34\%) \\
2 & $1 \times 10^6$ & 3971.13 & 18.90 & 6.50 & 7.05 (8.38\%) & 8.28 (27.40\%) \\
2 & $2 \times 10^6$ & 8504.98 & 26.17 & 8.28 & 8.88 (7.30\%) & 10.66 (28.79\%) \\
2 & $5 \times 10^6$ & 23296.22 & 40.00 & 11.25 & 12.12 (7.67\%) & 14.90 (32.39\%) \\
2 & $1 \times 10^7$ & 43690.91 & 54.01 & 14.11 & 14.85 (5.27\%) & 18.36 (30.14\%) \\
2 & $2 \times 10^7$ & 87605.10 & 72.54 & 17.57 & 18.52 (5.39\%) & 23.15 (31.73\%) \\
2 & $5 \times 10^7$ & 204899.56 & 107.28 & 23.41 & 24.49 (4.62\%) & 30.72 (31.25\%) \\
2 & $1 \times 10^8$ & 419021.34 & 141.89 & 29.07 & 30.72 (5.68\%) & 38.99 (34.13\%) \\
5 & $1 \times 10^6$ & 1215.97 & 8.56 & 6.10 & 6.74 (10.55\%) & 7.58 (24.38\%) \\
5 & $2 \times 10^6$ & 2424.42 & 12.39 & 7.82 & 8.42 (7.65\%) & 9.53 (21.84\%) \\
5 & $5 \times 10^6$ & 5834.48 & 21.25 & 10.76 & 11.35 (5.52\%) & 12.75 (18.54\%) \\
5 & $1 \times 10^7$ & 11543.18 & 28.71 & 13.61 & 14.12 (3.78\%) & 16.00 (17.58\%) \\
5 & $2 \times 10^7$ & 22446.05 & 39.75 & 17.08 & 17.59 (3.00\%) & 19.96 (16.84\%) \\
5 & $5 \times 10^7$ & 55357.57 & 60.42 & 23.01 & 23.71 (3.06\%) & 26.96 (17.17\%) \\
5 & $1 \times 10^8$ & 107172.05 & 79.49 & 28.64 & 29.40 (2.64\%) & 33.59 (17.29\%) \\
10 & $1 \times 10^6$ & 372.05 & 3.50 & 5.36 & 5.82 (8.45\%) & 6.45 (20.28\%) \\
10 & $2 \times 10^6$ & 835.21 & 6.08 & 7.22 & 7.65 (5.97\%) & 8.42 (16.67\%) \\
10 & $5 \times 10^6$ & 2062.50 & 11.98 & 10.09 & 10.54 (4.47\%) & 11.36 (12.66\%) \\
10 & $1 \times 10^7$ & 3955.05 & 17.98 & 12.72 & 13.18 (3.63\%) & 14.11 (10.92\%) \\
10 & $2 \times 10^7$ & 7875.63 & 23.67 & 16.05 & 16.43 (2.37\%) & 17.74 (10.55\%) \\
10 & $5 \times 10^7$ & 19137.64 & 39.09 & 21.71 & 22.29 (2.64\%) & 23.84 (9.80\%) \\
10 & $1 \times 10^8$ & 37057.30 & 50.27 & 27.13 & 27.59 (1.69\%) & 29.71 (9.52\%) \\
10 & $2 \times 10^8$ & 72836.27 & 67.87 & 33.89 & 34.54 (1.91\%) & 37.21 (9.80\%) \\
10 & $5 \times 10^8$ & 167572.38 & 98.02 & 45.20 & 45.61 (0.91\%) & 49.12 (8.66\%) \\
10 & $1 \times 10^9$ & 333601.74 & 129.57 & 55.91 & 57.27 (2.43\%) & 61.79 (10.52\%) \\
100 & $1 \times 10^7$ & 51.94 & 0.88 & 7.86 & 7.02 (10.63\%) & 7.20 (8.33\%) \\
100 & $2 \times 10^7$ & 151.22 & 2.05 & 10.81 & 10.19 (5.73\%) & 10.25 (5.18\%) \\
100 & $5 \times 10^7$ & 354.76 & 4.57 & 15.09 & 13.88 (8.01\%) & 13.61 (9.82\%) \\
100 & $1 \times 10^8$ & 784.51 & 7.00 & 19.16 & 18.12 (5.43\%) & 17.72 (7.55\%) \\
100 & $2 \times 10^8$ & 1550.46 & 11.36 & 24.35 & 22.97 (5.69\%) & 22.23 (8.72\%) \\
100 & $5 \times 10^8$ & 3338.95 & 18.75 & 32.32 & 29.91 (7.47\%) & 28.70 (11.21\%) \\
100 & $1 \times 10^9$ & 6491.31 & 26.64 & 40.14 & 37.44 (6.71\%) & 35.81 (10.78\%) \\
\end{tabular}
  \caption{Values of the parameters $A$ and $C_m$ of the eddy thermal diffusivity model. $Nu$ (model) are the $Nu$ values estimated using the eddy diffusivity model with these values of $A$ and $C_m$ and $1/[I_1(1/2)]$ are the values of $Nu$ obtained when $K(x)/\nu$ is modeled instead by $A(x/H)^3$. The numbers in parentheses are the relative errors of the estimated values of $Nu$ with respect to the DNS data by~\citet{HNVL2022}.} \label{tab:tab1}
  \end{center}
\end{table}

\begin{figure}
 \centerline{\includegraphics[scale=0.35]{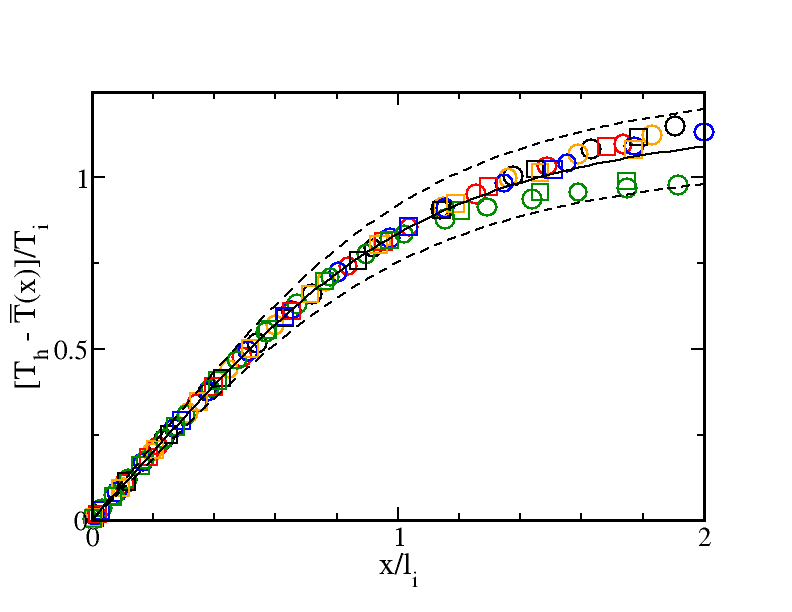}}
  \caption{$(T_h-\overline{T})/T_i$ vs $x/l_i$ at $Ra_{min}$~(circles) and  $Ra_{max}$~(squares) for $\Pran=1$~(black), $\Pran=2$~(red), $\Pran=5$~(orange),
   $\Pran=10$~(blue) and $\Pran=100$~(green). Here, $Ra_{min}$ and $Ra_{max}$ are the minimum and maximum values of $Ra$ for the corresponding $\Pran$~(see table~\ref{tab:tab1}). The solid line is the inner scaling function $F_i$ given by (\ref{scalinginner}) and the two dashed lines indicate $0.9F_i$ and $1.1F_i$.  }
\label{fig:Nfig2}
\end{figure}

In figure~\ref{fig:Nfig2}, we plot $(T_h-\overline{T} )/T_i$ as a function of $x/l_i$. The DNS data for different $\Pran$ and $Ra$ collapse onto a single curve that is well described by the theoretical inner scaling function $F_i$ in (\ref{scalinginner}) for $0 \le x/l_i < 2$, validating (\ref{Tinner}). In this region, $F_i$ clearly deviates from a linear function showing that  the turbulent heat flux $\overline{u'T'}$ cannot be neglected even within the inner region. 
Following~\citet{GC1979}, we express the temperature and length scales in the inner region on heat flux, or equivalently, $q$, the buoyancy parameter $\alpha g$ and the molecular diffusivities $\nu$ and $\kappa$. Dimensional analysis then yields
\begin{eqnarray} 
T_i =   \left( \frac{q^3}{\alpha g \kappa} \right)^{\frac{1}{4}} g_i(\Pran) = T_{i,GC} g_i(\Pran) = \Delta T Nu^{3/4} (Ra \Pran)^{-\frac{1}{4}} g_i(\Pran) 
\label{Ti} 
\end{eqnarray}
up to some $\Pran$-dependent function $g_i(\Pran)$. Using (\ref{qqrelation}) and (\ref{Vi}), we have
\begin{equation}
l_i =  \left( \frac{\kappa^3}{\alpha g q}\right)^{\frac{1}{4}} g_i(\Pran) = l_{i,GC} g_i(\Pran) 
\label{li}
\end{equation}

\begin{figure}[h]
  \centerline{\includegraphics[scale=0.35]{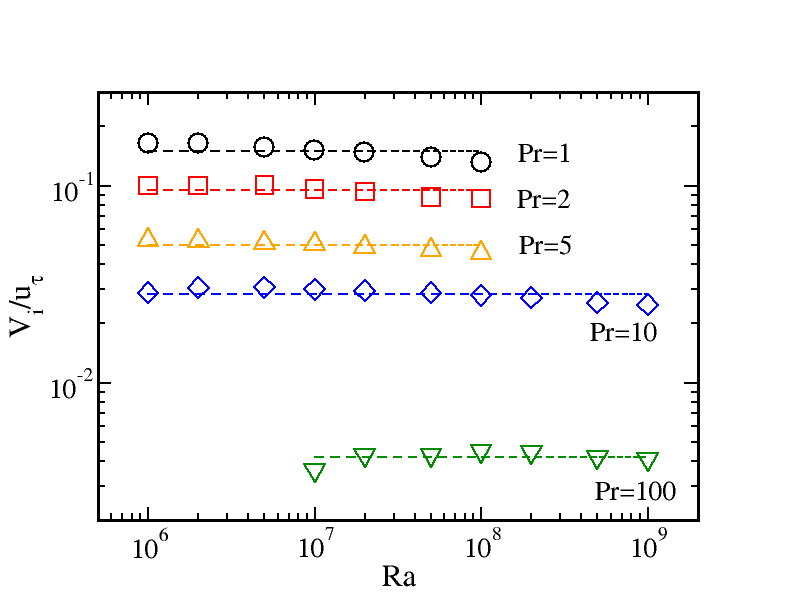}}
  \caption{ $V_i/u_\tau$ as a function of $Ra$. The dashed lines give the average values of $V_i/u_\tau$ over $Ra$ for each $\Pran$.} 
\label{fig:Nfig3}
\end{figure}

Our inner temperature and length scales $T_i$ and $l_i$ thus differ from the scales $T_{i,GC} = (q^3/\alpha g \kappa)^{1/4}$ and $l_{i,GC}=(\kappa^3/\alpha g q)^{1/4}$ adopted by~\citet{GC1979}. The inclusion of $\nu$ in our analysis leads to an additional function $g_i(\Pran)$, and this enables us to obtain a universal scaling function $F_i$ in the inner region for all $\Pran$. 
By comparing (\ref{Tiscale}) with (\ref{Ti}), we obtain
\begin{eqnarray}
(\Pran A)^{-1/3} = (Ra \Pran Nu)^{-1/4} g_i(\Pran)
\label{g}
\end{eqnarray}
which is used to evaluate $g_i(\Pran)$. We find that
 $g_i(\Pran)$ can be fitted by a power law $k_i \Pran^{\beta}$ with $\beta = 0.425 \approx 3/7$ and $k_i = 2.18$.  A commonly used velocity scale in the inner region is the friction velocity $u_\tau = \sqrt{\nu d\overline{w}/dx |_{x=0}}$. It is interesting to compare the inner velocity scale $V_i$ and $u_\tau$. Using (\ref{Vi}) and (\ref{g}), we obtain
 \begin{equation}
 \frac{V_i}{u_\tau}  =\frac{(Ra Nu)^{1/4} }{g_i(\Pran) \Pran^{3/4} \Rey_\tau}
 \label{Viutau}
 \end{equation} 
 where $\Rey_\tau=u_\tau H/\nu$ is the shear Reynolds number. The theoretical results of~\citet{Ching2023} that $Nu$ and $\Rey_\tau$ scale as $Ra^{1/3}$ in the high-$Ra$ limit then imply that $V_i$ is related to $u_\tau$ by a $\Pran$-dependent function $h(\Pran)$ when $Ra$ is sufficiently large:
 \begin{equation}
 V_i \approx h(\Pran) u_\tau \label{h}
 \end{equation}
It can be seen in figure~\ref{fig:Nfig3} that (\ref{h}) is confirmed and $h(\Pran)$ is a decreasing function of $\Pran$. As a result, if the mean temperature profiles are rescaled by $l_\tau \equiv \kappa/u_\tau$ and $T_\tau\equiv q/u_\tau$ instead of $l_i$ and $T_i$, the profiles for different $\Pran$ do not collapse beyond the linear region near the wall~[as shown in figure 5 b of~\citep{HNVL2022}].

\begin{figure}
  \centerline{\includegraphics[scale=0.35]{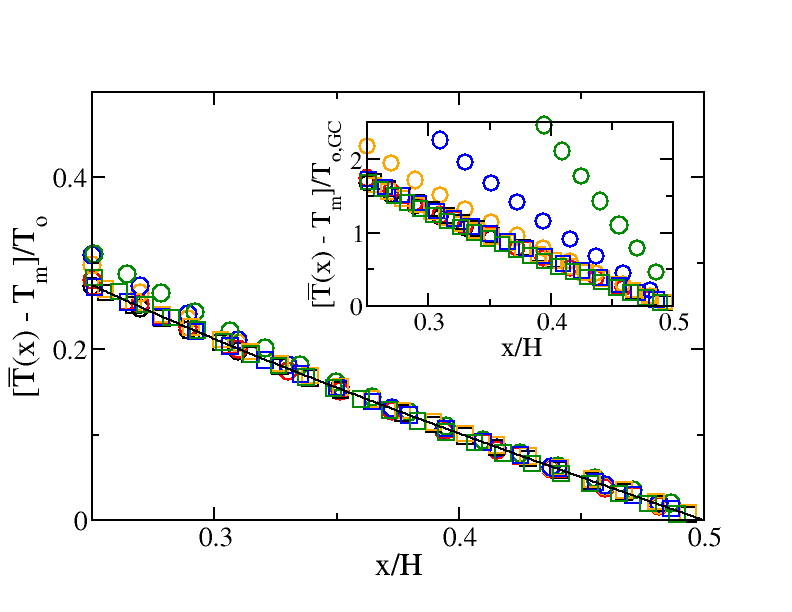}}
  \caption{Plot of $(\overline{T} - T_m)/T_o$ vs $x/H$ at $Ra_{min}$ and  $Ra_{max}$ for $\Pran=1, 2, 5, 10, 100$. Same symbols as in figure~\ref{fig:Nfig2}. The solid line is the  function $F_o$ given by  (\ref{scalingouter}). A similar plot is shown in the inset with the temperature rescaled by $T_{o,GC}$ instead of $T_o$. }
\label{fig:Nfig4}
\end{figure}

\begin{figure}
 \centerline{\includegraphics[scale=0.4]{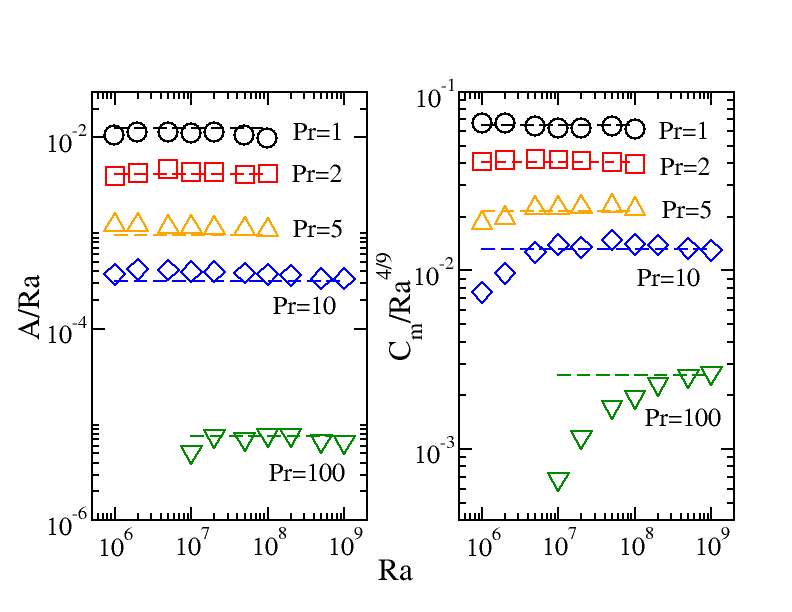}}
 \caption{Plot of $A/Ra$ and $C_m/Ra^{4/9}$ vs $Ra$.The dashed lines for different $\Pran$ are the values of $C_1(\Pran)$ and $C_2(\Pran)$ obtained by using $C=0.043$, $f(\Pran)\approx 0.19$~\citep{Ching2023} and the fitted values of $k_i$, $\beta$ and $k_o$~[see (\ref{Adep}) and (\ref{Cmdep})].
\protect\\    }
\label{fig:Nfig5}
\end{figure}

Next we check (\ref{Touter}). As shown in figure~\ref{fig:Nfig4}, the rescaling of $\overline{T}-T_m$ by $T_o$ results in an approximate collapse of the data in the region $0.3 \le x/H \le 0.5$. The collapsed data are consistent with the outer scaling function $F_o$ with $b=4$ in (\ref{scalingouter}).  These observations support our choice of $y_2=0.3$ and $b=4$ and validate our proposed functional form of $K(x)$ in the outer region.  If the outer temperature scale $T_{o,GC}$ proposed by~\citet{GC1979} is used instead of $T_o$, the data collapse is worse~(see the inset of figure~\ref{fig:Nfig5}). The scale $T_{o,GC}$, given by
\begin{equation}
T_{o,GC}=\left(\frac{q^2}{\alpha g H} \right)^{1/3} ,
\label{ToGC}
\end{equation}
was obtained by assuming that the  temperature scale in the outer region depends on $q$, $\alpha g$ and $H$ only and not on the molecular diffusivities $\nu$ and $\kappa$. We find that  
\begin{equation}
C_m \to k_o (Ra Nu)^{1/3} \Pran^{-2/3} 
\label{Cm}
\end{equation}
with $k_o = 0.162$ for sufficiently large $Ra$. 
This implies that $T_o \to k_o^{-1} T_{o,GC}$ when $Ra$ is greater than certain threshold value $Ra_c(\Pran)$ and $Ra_c(\Pran)$ increases with $\Pran$. 

Finally, we show that our model gives $Nu \sim Ra^{1/3}$ in the high-$Ra$ limit.
Using (\ref{g}) and (\ref{Cm}), it can be shown that for sufficiently large $Ra$, $I_1(y_1) \sim  (Ra Nu)^{-1/4}$, $I_2(y_2) \sim (Ra  Nu)^{-1/3} \ln (Ra  Nu)$ and 
 $I_3(y_2) \sim (Ra Nu)^{-1/3}$ and hence
\begin{eqnarray}
 Nu  \approx \frac{1}{2I_1(y_1)} 
  \Rightarrow Nu \sim Ra^{1/3} .
 \label{NuModel}
\end{eqnarray}
Substitute the theoretical result of $Nu \approx [C^2f(\Pran)]^{1/3}\Pran^{-1/9} Ra^{1/3} = \gamma(\Pran) Ra^{1/3}$ for $\Pran \ge 1$ in the high-$Ra$ limit~\citep{Ching2023} into (\ref{g}) and (\ref{Cm}), we obtain 
\begin{eqnarray}
A &\approx& [\gamma(\Pran)]^{3/4} [g_i(\Pran)]^{-3} \Pran^{-1/4} Ra = C_1(\Pran) Ra \\
\label{Adep}
C_m &\approx& k_o [\gamma(\Pran)]^{1/3} \Pran^{-2/3} Ra^{4/9} = C_2(\Pran) Ra^{4/9}
\label{Cmdep}
\end{eqnarray}
when $Ra$ is sufficiently large. Such $Ra$-dependencies of $A$ and $C_m$ are confirmed in figure~\ref{fig:Nfig5}.

\subsection{Comparison with mean temperature profiles reported in previous studies}

\begin{figure}
  \centerline{\includegraphics[scale=0.4]{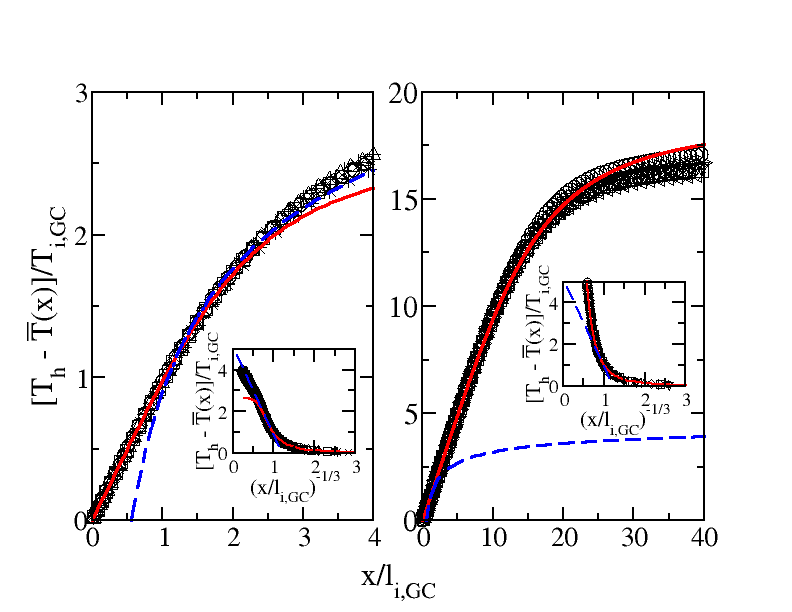}}
  \caption{Plots of $[T_h - \overline{T}(x)]/T_{i,GC}$ vs $x/l_{i,GC}$ for $\Pran=1$ (left panel) and $\Pran=100$ (right panel) at $Ra = 10^6$~(plusses), $2\times 10^6$~(crosses), $5\times10^6$~(stars), $10^7$~(circles), $2\times 10^7$~(squares), $5\times 10^7$~(diamonds), $10^8$~(triangles), $2\times 10^8$~(left triangles), $5\times 10^8$~(inverted triangles) and $10^9$~(right triangles). The red solid lines are (\ref{Tinner2}) and the blue dashed lines are (\ref{inverseCubicIn}) with $K_1=4.2$, $\phi_1(1)=5.10$ and $\phi_1(100)=5.15$.  
}
\label{fig:Nfig6}
\end{figure}

One often-cited result for mean temperature profile was obtained by~\citet{GC1979}.  
They proposed scaling functions in terms of $T_{i,GC}$ and $l_{i,GC}$ in the inner region and $T_{o,GC}$ and $H$ in the outer region. In an overlap layer in which both scaling functions hold, they obtained
\begin{eqnarray}
\frac{T_h - \overline{T}(x)}{T_{i,GC}} &=& - K_1 \left(\frac{x}{l_{i,GC}}\right)^{-1/3} + \phi_1(\Pran) \label{inverseCubicIn}, \\
 \frac{\overline{T}(x)-T_m}{T_{o,GC}} &=& K_1 \left(\frac{x}{H}\right)^{-1/3} + \theta_1
\label{inverseCubicOut}
\end{eqnarray}
with undetermined constants $K_1$, $\phi_1(\Pran)$ and $\theta_1$; the independence of $K_1$ and $\theta_1$ on $\Pran$ follows from the independence of the outer scaling function on $\nu$ or $\kappa$. Fitting their DNS data for $\Pran=0.709$~(air), \citet{VN1999} and \citet{NCO2013} found $K_1=4.2$. Using the DNS data of~\citet{HNVL2022}, we find that the region that (\ref{inverseCubicIn}) with $K_1=4.2$ can fit decreases drastically for larger $\Pran$  while our result for the inner region
 (\ref{Tinner}) with (\ref{scalinginner}), rewritten as
\begin{eqnarray}
\frac{T_h - \overline{T}(x)}{T_{i,GC}} = g_i(\Pran) F_i \left(\frac{1}{g_i(\Pran)}\frac{x}{l_{i,GC}}\right) ,
\label{Tinner2}
\end{eqnarray} can give good fits for all the 5 values of $\Pran$ studied. The comparison for $\Pran=1$ and $\Pran=100$ is shown in figure \ref{fig:Nfig6}. 
 
\begin{figure}
 \centerline{\includegraphics[scale=0.4]{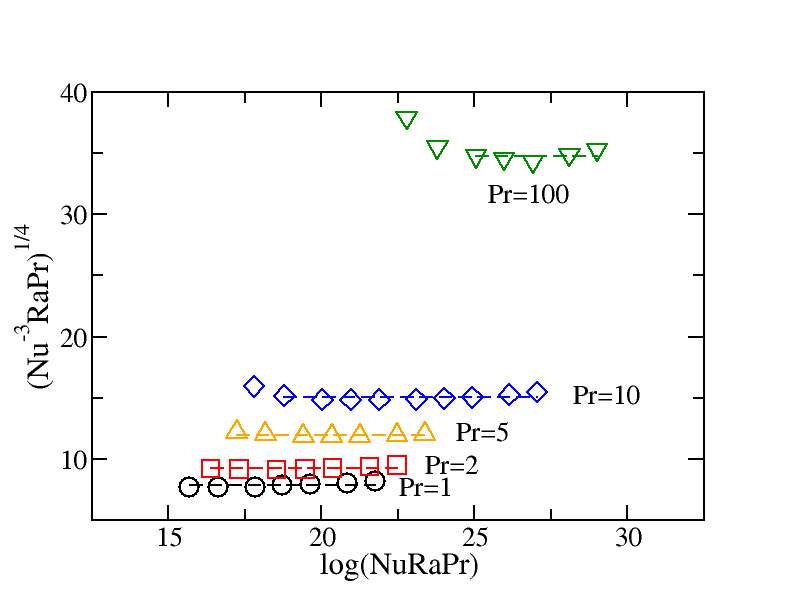}}
 \caption{Plot of $(Nu^{-3} Ra \Pran)^{1/4}$ vs  $\log(Nu Ra \Pran)$ using DNS data of~\citet{HNVL2022}. The dashed lines for different $\Pran$ are the best fits of (\ref{NuHH2005}) with $K_2=0$. }
\label{fig:Nfig7}
\end{figure}

Using the same approach but with the temperature scale $T_{i,GC}$ for both the inner and outer regions, \citet{HH2005} 
obtained a logarithmic profile in the overlap layer:
\begin{eqnarray}
\frac{T_h - \overline{T}(x)}{T_{i,GC}} &=& K_2 \log \left(\frac{x}{l_{i,GC}}\right) + \phi_2(\Pran)  \label{logIn} \\
 \frac{\overline{T}(x)-T_m}{T_{i,GC}} &=& - K_2 \log \left(\frac{x}{H}\right) + \theta_2
 \label{logOut}
\end{eqnarray}
with undetermined constants $K_2$, $\phi_2(\Pran)$ and $\theta_2$. Different values of $K_2$ and $\phi_2$, obtained by fitting DNS data for air, were reported~\citep{VN1999,HH2005,KH2012,NCO2013}. We thus test instead the relation between $Nu$, $Ra$ and $\Pran$, obtained by adding (\ref{logIn}) and (\ref{logOut}) and using
(\ref{Ti}) and (\ref{li}):
\begin{eqnarray}
(Nu^{-3} Ra \Pran)^{1/4}  = \frac{K_2}{2} \log(Nu Ra \Pran) + 2[\phi_2(\Pran)+\theta_2]
\label{NuHH2005}
\end{eqnarray}
\citet{BHH2007} assumed (\ref{logIn}) to hold up to $x=H/2$ and obtained a different relation than (\ref{NuHH2005}).
As shown in figure~\ref{fig:Nfig7}, the DNS data~\citep{HNVL2022} are consistent with (\ref{NuHH2005}) with $K_2=0$ indicating their incompatibility with (\ref{logIn}) and (\ref{logOut}). When $K_2=0$, (\ref{NuHH2005}) reduces to $Nu \sim Ra^{1/3}$.
 
\begin{figure}
 \centerline{\includegraphics[scale=0.4]{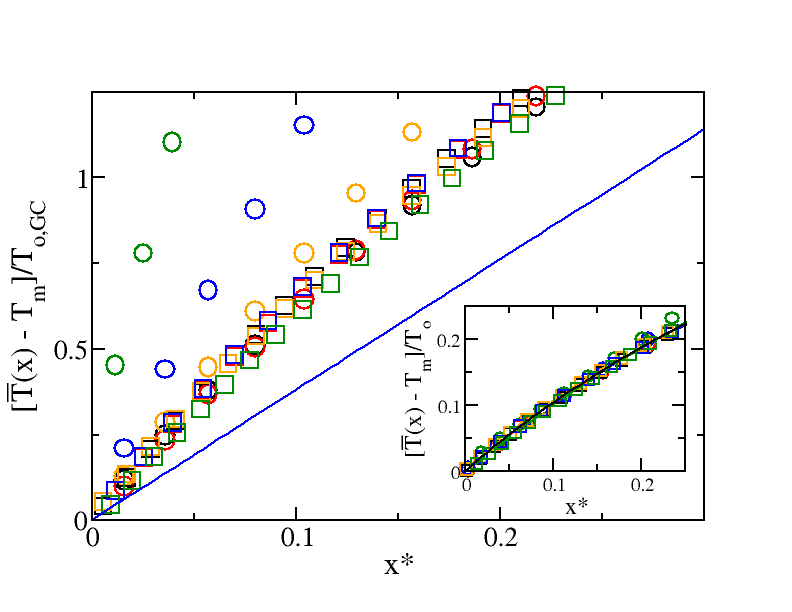}}
 \caption{Plot of $(\overline{T} - T_m)/(T_{o,GC})$ vs $x^*$ at $Ra_{min}$ and  $Ra_{max}$ for $\Pran=1, 2, 5, 10, 100$. Same symbols as in figure~\ref{fig:Nfig2} and the solid straight line is $c x^*$. In the inset, a similar plot is shown with the temperature rescaled by $T_{o}$ instead of $T_{o,GC}$ and the solid line is (\ref{Touter2}).}
\label{fig:Nfig8}
\end{figure}

\vspace{0.5cm}
A recent work~\citep{LJLJZ2023} studied the mean velocity and temperature profiles by modeling the Reynolds stress and turbulent heat flux:
\begin{eqnarray} 
\overline{u'w'} &=& -k_1 x^2 \left( \frac{\partial \overline{w}}{\partial x} \right)^2 \label{Li1} \\
\overline{u'T'} &=& -k_2 x^2 \frac{\partial \overline{w}}{\partial x} \frac{\partial \overline{T}}{\partial x} \label{Li2}
\end{eqnarray}
where $k_1$ and $k_2$ are dimensionless coefficients.
As discussed in Sec.~(\ref{Emodel}), $\overline{u'T'}$ and its first- and second-order derivatives with respect to $x$ vanish at $x=0$ due to the boundary conditions. Similar arguments require 
$\overline{u'w'}$ and its first- and second-order derivatives with respect to $x$ to vanish at $x=0$. These properties should be satisfied by any 
closure model but they are violated by (\ref{Li1}) and (\ref{Li2}). Since both the mean velocity and temperature gradients, $\partial \overline{w}/\partial x$ and $\partial \overline{T}/\partial x$, are non-zero at $x=0$, the model expressions on the RHS of (\ref{Li1}) and (\ref{Li2}) have non-vanishing second-order derivatives at $x=0$. Applying (\ref{Li1}) and (\ref{Li2}) to the near-centerline outer region, Li et al. obtained an inverse cubic-root dependence for the mean temperature profile~[(2.14b) of~\citep{LJLJZ2023}], which we rewrite as:
\begin{equation}
\frac{\overline{T}(x)-T_m}{T_\tau} = c K\left[  \left(\frac{x}{H} \right)^{-1/3} - 2^{1/3} \right]  =c K x^*
\label{LiT}
\end{equation}
where $K = (\alpha g H T_\tau/u_\tau^2)^{1/3}$, $c \approx 3.8$ and
\begin{equation}
x^*\equiv  \left(\frac{x}{H} \right)^{-1/3} - 2^{1/3}
\label{x*}
\end{equation}
We note that (\ref{LiT}) implies that the mean temperature profiles in the outer region would collapse onto a straight line in $x^*$ when temperature is rescaled by $T_\tau/K=T_{o,GC}$. 
Using (\ref{x*}), we rewrite our result for the outer region, (\ref{Touter}) with (\ref{scalingouter}), in terms of $x^*$:
\begin{equation}
\frac{\overline{T}(x)-T_m}{T_o} = \frac{1}{4} \log[(x^*+2^{1/3})^3 - 1]
\label{Touter2}
\end{equation}
In figure~\ref{fig:Nfig8}, it can be clearly seen that the mean temperature profiles in the outer region do not collapse onto the straight line $c x^*$ when rescaled by $T_{o,GC}$, contrary to the prediction by (\ref{LiT}), but they are well represented by (\ref{Touter2}).

\section{Conclusions}

In this study, we focus to determine the mean temperature profiles in turbulent natural convection between two infinite vertical walls at different temperatures.
A three-layer model for the eddy thermal diffusivity with two parameters has been proposed. Using this model, we have derived analytical results for the mean temperature profiles in terms of the two parameters of the model. These analytical results reveal that the mean temperature profiles for different $Ra$ and $\Pran$ are described by two universal scaling functions in the inner region next to the walls and the outer region near the centerline between the walls. The characteristic temperature scales in these two regions are expressed in terms of the two parameters, which determine the characteristic velocities for heat transfer in the two regions. The dependencies of the two parameters on $Ra$ and $\Pran$ are obtained. Analytical expressions of  the two scaling functions are fully determined and there is no overlap region in which both functions hold.   Our theoretical results for the mean temperature profiles in the inner and outer regions are in good agreement with DNS data for $1 \le \Pran \le 100$ and $10^6 \le Ra \le 10^9$~\citep{HNVL2022} and can give a better description of the data than results reported in the literature~\citep{GC1979,HH2005,LJLJZ2023}.

\begin{bmhead}[Funding.]
This work was funded by the Hong Kong Research Grants Council (Grant No. CUHK 14303623).
\end{bmhead}

\begin{bmhead}[Declaration of interests.]
The authors report no conflicit of interest.
\end{bmhead}

\begin{bmhead}[Author ORCIDs.]\\
Ho Yin Ng, https://orcid.org/0009-0003-8811-2743; \\
Emily S.C. Ching, https://orcid.org/0000-0001-5114-5072 
\end{bmhead}

\end{document}